\documentclass[a4paper,11pt]{article}
\usepackage[utf8]{inputenc}
\title{ Fidelity susceptibility for  Lifshitz geometries via  Lifshitz Holography  }
\author{Davood Momeni\\Department of Physics, College of Science, Sultan Qaboos University,\\
	P.O. Box 36, P.C. 123, Al-Khowd, Muscat, Sultanate of Oman\\\\
Mir Faizal\\ Irving K. Barber School of Arts and Sciences, 
\\ University of British
Columbia - Okanagan,  
3333 University Way,\\  Kelowna,   British Columbia V1V 1V7, Canada
\\ Department of Physics and Astronomy, University of Lethbridge,\\
Lethbridge, Alberta, T1K 3M4, Canada\\\\
Aizhan Myrzakul, Ratbay Myrzakulov\\ Eurasian International Center for Theoretical Physics \\
and Department of General Theoretical Physics, \\
Eurasian National University, Astana 010008, Kazakhstan
}
\date{}
\begin{document}

\maketitle

\begin{abstract}
In order to analyze the fidelity susceptibility of non-relativistic field theories, which 
are important in condensed matter systems, we generalize the proposal to obtain 
the fidelity susceptibility holographically to Lifshitz geometries. 
It will be argued that this proposal can be used to study 
 the fidelity susceptibility for various condensed matter systems. 
To demonstrated this, we will explicitly  use this proposal to analyze the  fidelity susceptibility 
for a non-relativistic many-body system, and argue  
  that the fidelity susceptibility of this theory can be holographically 
obtained from a bulk Lifshitz geometry. In fact, using a 
Einstein-Dilaton-Maxwell-AdS-Lifshitz theory, we explicitly 
  demonstrated that the 
 fidelity susceptibility obtained from this  
  bulk geometry is equal to the   fidelity susceptibility of a  
  bosonic many-body system. 
\end{abstract}
\section{Introduction}
 It is known that  the entropy of a black hole 
 scales with its area. As black holes are maximum entropy objects, this  implies that 
 the maximum entropy of that certain  region of space scales with the area of its  
 boundary. This observation has led to the development of the  holographic principle, which 
 equates the   number of degrees of
freedom in a region of space   to the number of degrees 
of freedom on the boundary surrounding that region of space \cite{1, 2}.
The     AdS/CFT correspondence  is a realization of the holographic principle as 
it is a duality between 
the string theory/supergravity in AdS spacetime 
and  the field theory on its boundary   \cite{M:1997}. 
As AdS/CFT correspondence is a duality between two very different theories, it 
seems from the AdS/CFT correspondence and the holographic principle that laws 
of physics are fundamentally just information theoretical processes. 
In fact, various   studies done in 
  different fields of science  seem to  indicate that the  laws of physics 
are informational theoretical processes \cite{info, info2}. 
So, the  AdS/CFT can   be used to obtain   information theoretical information 
relating to a   conformal field theory from the bulk geometry. 
The entanglement entropy of a field theory is a   most important 
informational theoretical quantity relating to a conformal field theory. 
It has been demonstrated that the entanglement entropy of a conformal field theory
can be holographically obtained from the bulk AdS spacetime, as it is dual a minimal surface in 
  asymptotically AdS spacetime
\cite{6, 6a}.  

It is also important to know how much information is retained in a system, 
and holographic entanglement entropy can be used to quantify this   as 
it measures the loss of information in a system. 
However, it is also important to know the difficulty to obtain this 
information, and this can be quantified using complexity. 
As laws of physics can be understood in terms of information 
theoretical processes \cite{info, info2}, and complexity is an important 
informational theoretical quantity, complexity is expected to be an important 
physical quantity used in the laws of physics. 
In fact, complexity has been used to understand the behavior of  
condensed matter systems \cite{c1, c2}
 and  molecular physics \cite{comp1},  quantum computing \cite{comp2}. 
 In fact, it has been argued that the information might not be ideally lost in a black hole, 
 but it would be effectively lost, as it would not be possible   obtain this 
 information from a black hole due to its chaotic nature   \cite{hawk}. 
 
 The complexity of a conformal field theory can also be obtained holographically, 
as the holographic complexity is dual to a volume in AdS spacetime 
 \cite{Susskind:2014rva1,Susskind:2014rva2,Stanford:2014jda,Momeni:2016ekm, Alishahiha:2015rta, Alishahiha:2017cuk}. 
It has been demonstrated that the holographic complexity of a field theory can 
 be related to the fidelity susceptibility of the boundary field theory. So,  the fidelity susceptibility of a field theory 
 can be holographically calculated using a  maximal volume  $ V(\gamma_{max})$ in the AdS which
ends on the time slice at the AdS boundary \cite{r5, r7}.  

As fidelity susceptibility is important to understand the behavior of condensed matter systems 
\cite{f1, f2,f4,f5,f7}, 
it is important to generalize this proposal to non-relativistic field theories describing 
condensed matter systems. It may be noted that such non-relativistic   
condensed matter systems can be holographically analyzed using Lifshitz holography 
\cite{cond1, cond2, cond4, cond5, cond6}. 
In the Lifshitz holography, the Lifshitz deformed of AdS can be related to the Lifshitz field theories, 
in which spacetime have different scaling behavior  \cite{Griffin:2012qx}.  As   
Schr\"{o}dinger invariant quantum system, which describe condensed matter system, the space and time scale differently,
we will 
use Lifshitz holography to analyze such a system.  It may be noted that a relation between the Lifshitz holography
and Schr\"{o}dinger invariant quantum system has been studied  for certain system \cite{0sh1, 0sh2}. As we will be studying a
  static case,  we will be able to find a holographic  relation between   a 
 Lifshitz system  and a Schr\"{o}dinger invariant quantum system.

\section{Lifshitz  Fidelity Susceptibility}
In this section,   motivating by  \cite{Griffin:2012qx}, in this work we will now generalize the relativistic  proposal 
to obtain  the fidelity susceptibility   from holographic complexity \cite{r5, r7}
to Lifshitz geometries. So, if we consider one parameter a Lifshitz field theory, with states denoted by $|\Psi(\lambda)>$, then 
the inner product of two such states separated by an infinitesimally perturbation $\delta \lambda$,  can be expressed as 
\begin{equation}
 <\Psi(\lambda|\Psi(\lambda +\delta\lambda)> = 1-\Xi_F (\delta \lambda)^2. 
\end{equation}
Now using the argument used in \cite{r5, r7}, it can be argued that $\Xi_F$ can be written as 
\begin{equation}
 \Xi_F = \int dm <O(x), O(x')>, 
\end{equation}
where $dm$ is a suitable integral measure for this system, and $O(x), O(x')$ are suitable local operator in the 
Lifshitz theory. Now it can again be argued that this quantity will be described holographically. 
However, as these operators are defined in a Lifshitz theory, and the holographic dual of such a field theory 
is described by Lifshitz gravity   \cite{Griffin:2012qx}, we can use the argument of \cite{r5, r7}, to argue that 
$\Xi_F$ can also be obtained holographically from a Lifshitz-AdS geometry. In fact, 
as this quantity has to reduce to holographic complexity for $z=1$ for usual theory, we can also calculate this quantity 
in Lifshitz geometries using the volume of maximum surfaces. 

This can be done by  defining $\mathcal{V}(\gamma_{max})$  as a 
  maximal volume in the Lifshitz deformation of the AdS spacetime, which
ends on the time slice at the Lifshitz-AdS boundary. 
We can use  this  
maximal volume in the Lifshitz geometry to define the  holographic complexity in such a geometry as 
\begin{eqnarray}
&&F=\frac{V(\gamma_{max})}{8\pi R G}, 
 \end{eqnarray}
 It may be noted that for $z=1$ this maximal volume reduces to the 
usual maximal volume in AdS spacetime, and so this  holographic complexity reduces to the usual
holographic complexity for $z=1$  \cite{r5, r7}.
Now   the Lifshitz holography reduces to the 
usual holography for $z =1$, and it is known that there are divergences associated with 
such volumes for $z=1$ 
\cite{Carmi:2016wjl}.
So, we need to regularize this volume, before we can define the fidelity susceptibility  
for Lifshitz geometries. This will be done by subtracting the background Lifshitz-AdS geometry ${V}(\gamma_{max})_{LAdS}$
from the deformed 
Lifshitz-AdS geometry ${V}(\gamma_{max})_{DLAdS}$. So, we can define a regularized Lifshitz maximal
volume  
\begin{eqnarray}
\mathcal{V}(\gamma_{max}) ={V}(\gamma_{max})_{DLAdS}-  {V}(\gamma_{max})_{LAdS}
 \end{eqnarray}
 Now using this  regularized 
maximal volume in the Lifshitz geometry, we can define the  regularized holographic complexity of a
non-relativistic boundary theory as 
\begin{eqnarray}
\Xi_F &=& F_{DLADS}- F_{LAdS}\nonumber \\ &= & \frac{\mathcal{V}(\gamma_{max})}{8\pi R G}, \label{b}
 \end{eqnarray}
where $R$ is the radius of the curvature of this  
Lifshitz-AdS geometry. This regularized holographic complexity is equal to fidelity susceptibility of the boundary field theory, 
and so it fidelity susceptibility of the boundary field theory can be holographically calculated from holographic complexity. 
It may be noted that for $z=1$ this expression  reduces to the 
usual expression for the  regularized 
fidelity susceptibility  \cite{rf14, rf17}. 
\section{Boundary Bosonic System}
Now we  will  use this proposal for holographically analyze a simple system of   
system of $N$ bosons, without any self-interaction. These bosons will be placed   
in a background  uniform magnetic 
field $\vec{H}$ in the $z$ direction, with  $\vec{H}=H\hat{e}_z$. So, even though the bosons do not interact 
with a dynamical field, they do interact with a  background field. 
It is possible to holographically analyze such   systems 
in a background field \cite{b1a0,b2a0,b4a0, b6a0}. We will analyze this proposal for such a simple system, 
to demonstrate how such a holographic correspondence can work, and so this proposal can be used for holographically analyzing 
more complicated systems. 
Now before we analyze the bulk Lifshitz geometry dual to such a system,
we will calculate the fidelity 
susceptibility of this bosonic  theory. 
The Hamiltonian for each of these bosonic particles is  
$
 H_i= {(-i\vec{\nabla}_i-q\vec{A})^2}/{2m} ,
$
where $\vec{A}=\vec{\nabla}\times \vec{H}$, so the 
time-dependent Schr\"{o}dinger equation for these bosonic particles can be written as 
 \begin{eqnarray}
 \sum\frac{(-i\vec{\nabla}_i-q\vec{A})^2}{2m}\Psi_{tot}(\vec{x}_1,\vec{x}_2,...,\vec{x}_N;t)
 =i\frac{\partial}{\partial t}\Psi_{tot}(\vec{x}_1,\vec{x}_2,...,\vec{x}_N;t). &&
\end{eqnarray}

 Now we can write the  total wave function as 
 \begin{equation}
\Psi_{tot}(\vec{x}_1,\vec{x}_2,...,\vec{x}_N;t)=\Pi_{i=1}^N\psi_{i}(\vec{x}_i;t). 
 \end{equation}
We need to find only ground state wave function 
$\Psi^{0}_{tot}(\vec{x}_1,\vec{x}_2,...,\vec{x}_N;t)$. 
Choosing the  gauge,   $\vec{A}=(0,Hx,0)$, we can write 
$\vec{A}=\vec{\nabla}\times \vec{H}$. 
Using this gauge for $\vec{A}$, Schr\"{o}dinger equation can be expressed as 
 \begin{eqnarray}
&&-\nabla_i^2\psi_{i}(\vec{x}_i;t)+2iqHx\frac{\partial \psi_{i}
(\vec{x}_i;t)}{\partial y} +q^2H^2x^2\psi_{i}(\vec{x}_i;t)=2mE_i\psi_{i}(\vec{x}_i;t)\label{H2}.
\end{eqnarray}
We can express  
$\psi_i(\vec{x}_i;t)$ as 
\begin{equation}
 \psi_{i}(\vec{x}_i;t)=e^{i(k_i z_i+\beta_i y_i-E_i t)}\phi_{i}(x_i).
\end{equation}
Now we can write 
\begin{eqnarray}
 -\frac{d^2\phi_{i}(x_i)}{dx_i^2}+\Big(q^2H^2x_i^2-2q\beta_iH x_i\Big)\phi_{i}(x_i)
 =(2mE_i-k_i^2-\beta_i^2)\phi_{i}(x_i)\label{1D-phi}.&&
\end{eqnarray}
It may be noted that this looks just like the Schr\"{o}dinger equation for 
a simple harmonic oscillator, such that the coordinates have been  shifted
as $x_i\to \xi+\frac{\beta_i}{qH} $. So, we can express the Schr\"{o}dinger equation 
 for this system as 
\begin{eqnarray}
-\frac{1}{2m}\frac{d^2\phi_{i}(\xi_i)}{d\xi_i^2}+\frac{1}{2}m\omega_i^2\xi_i^2\phi_{i}(\xi_i)=\Big(E_i-\frac{k_i^2}{2m}\Big)\phi_{i}(\xi_i)
\label{1D-psi},
\end{eqnarray}
where $\omega_i=\frac{qH}{m}$ denotes frequency (energy) of the system. 
Exact solution for  this equation can be expressed  in terms 
of Hermite functions. Now using the  
  ground state wave function for a bosonic particle,  
$ 
\phi_{i,0}(\xi_i)=\sqrt[4]{\frac{qH}{\pi}}e^{-\frac{qH}{2}\xi_i^2},\ \ E_{i,0}
=\frac{qH+k_i^2}{2m}.
$, the  ground state wave function for whole system can be written as 
 \begin{eqnarray}
\Psi_{tot,0}(\vec{x}_1,\vec{x}_2,...,\vec{x}_N;t)&=&(\frac{qH}{\pi})^{N/4}\Pi_{i=1}^N 
\nonumber \\ && \times e^{i\Big(k_i z_i+\beta_i y_i-(\frac{qH+k_i^2}{2m}) t\Big)-\frac{qH}{2}(x_i-\frac{\beta_i}{qH})^2}.
 \end{eqnarray}
 The  fidelity susceptibility can be obtained by varying  $H\to H+\delta H$, 
and computing the following inner product, 
\begin{equation}
F= <{\Psi_{tot,0}(H)|\Psi_{tot,0}(H+\delta H)}>= 1-(\delta H)^2\Xi_F+...
\end{equation}
So,  we expand  $F$ in series up to the second order \cite{b1}, 
 \begin{eqnarray}
&&<{\Psi_{tot,0}(H)|\Psi_{tot,0}(H+\delta H)}>
\nonumber \\ &=& \Pi_{i=1}^N\int d^3x_i \Psi^{\star}_{tot,0}(H+\delta H)\Psi_{tot,0}(H)
\\\nonumber &=& \Big(\int d^3x\psi^{\star}(\vec{x};t,H+\delta H)\psi(\vec{x};t,H)
\Big)^N. 
\end{eqnarray}
Thus, we can express the  fidelity susceptibility of this system as 
\begin{eqnarray}
&&\Xi_F=\frac{N}{8 q H^3}(qH+4\beta^2)\label{XIF}.
 \end{eqnarray}
Expression given in (\ref{XIF}) is the exact fidelity susceptibility for
a system of  $N$ charged bosonic  particles in a uniform magnetic field.  

\section{Lifshitz Bulk Dual}
This  boundary theory  studied in the previous section is defined by non-relativistic Hamiltonian. 
Now we will study a theory with Lifshitz symmetry   in the
bulk theory \cite{a1, a2, a4, a5}.
Now we will suppose that the Einstein-Dilaton-Maxwell-AdS-Lifshitz 
bulk action   \cite{k1, k2, l1, l2}. 
 We would like to point out that the  Schr\"{o}dinger  symmetry is  
different from    Lifshitz symmetry. However, a relation between the Lifshitz holography 
and certain Schr\"{o}dinger invariant quantum  systems has been studied \cite{0sh1, 0sh2}.  
Thus, it will be interesting to note that we will demonstrate that the 
fidelity  susceptibility  obtained from this Lifshitz theory will match the results calculated in the previous section 
for the theory with Schr\"{o}dinger  symmetry. 
This occurs as we   study a 
  static case, and so, it  would be interesting to analyze if 
  such a  match occurs for other holographic calculations 
      between such theories, for a static case.  
Thus, we propose the following action for the bulk theory \cite{l1, l2}, 
\begin{eqnarray}
S_{\mbox{Bulk}}=\int_{\mathcal{M}} d^{4}x\sqrt{-g}\left[\frac{(R-2\Lambda)}{2\kappa^2}-\frac{1}{2}
\partial^\mu\phi\partial_\mu\phi
+V(\phi)
-\xi\mathrm{e}^{\lambda\phi}(F^{\mu\nu}F_{\mu\nu})\right]\,.\label{action}
\end{eqnarray}
where   the potential is  $V(\phi)=V_0\mathrm{e}^{\gamma\phi}$ with parameters  
$V_0$ and $\gamma$. Here $\phi$ 
is non-minimally coupled with electromagnetic 
potential, and the 
electromagnetic field strength coupled to scalar field as
$\xi\mathrm{e}^{\lambda\phi}(F^{\mu\nu}F_{\mu\nu})$, such that  $\xi, 
\lambda$ are suitable  constants.  

The metric for a 
static, spherically symmetric solution in this Einstein-Dilaton-Maxwell-AdS-Lifshitz 
can be written as \cite{l1, l2}
\begin{equation}
ds^2=-e^{2\alpha(r)}B(r)dt^2+\frac{d r^2}{B(r)}+r^2d\sigma_{2,k}^2
\,,\label{metric}
\end{equation}
where $\alpha(r), B(r)$ are function of $r$. Here  $d \sigma_{2,k}^2$
is the metric for a topological two-dimensional surface parametrized 
by $k= 0, \pm 1$. This two-dimensional   manifold is a sphere $S_{2}$ for $k=1$, 
a torus $T_{2}$ for $k=0$, and a compact hyperbolic manifold $Y_{2}$ for $k=-1$.  
Now  we can choose $k=0$ and write    the  planar Euclidean coordinates as 
$d\sigma_{2,k}^2=dx^idx_i,\ \ i=1,2,x^i=\{x,y\}$.
 It may be noted that due to  Lifshitz scaling, 
$\alpha(r)\propto\log r^{z/2}$, where  $z$ is the Lifshitz parameter. 
So, the  general form of the metric  with Lifshitz symmetry can be written as  
\begin{equation}
ds^2=-\left(\frac{r}{r_0}\right)^{z}B(r)dt^2+\frac{d r^2}{B(r)}+r^2 dx_i dx^i\,.\label{exsolution}
\end{equation}
Here the function $B(r)$ can be written as \cite{l1, l2} 
\begin{eqnarray}
 B(r)&=&\frac{2}{(2+z)}\left[\frac{\tilde V_0}{2}\right]+ 
\Big(\frac{r_{+}}{r}\Big)^{1+z/2}\Big(\Big(\frac{2(\Lambda+\tilde Q^2\xi)}{(6+z)}\Big)r_{+}^2
\nonumber \\ && -\frac{2}{(2+z)}\left[\frac{\tilde V_0}{2}\right]
\Big) -\Big(\frac{2(\Lambda+\tilde Q^2\xi)}{(6+z)}\Big)r^2. 
 \end{eqnarray}
This bulk theory is dual to the bosonic system, we have analyzed in the previous section. 

\section{Holographic Complexity}
As we have obtained the fidelity susceptibility of the bosonic system in the previous section, we will 
holographically analyze it in this section. So, 
 we will use  generalization the 
   fidelity susceptibility \cite{r5, r7} to a Lifshitz geometry given by Eq. (\ref{b}). The fidelity 
   susceptibility in such geometries depends on 
   $V(\gamma_{max})$, and 
  we can  
obtain $V(\gamma_{max})$ using  
\begin{eqnarray}
&&V(\gamma_{max})=\int_{r_{+}}^{r_{\infty}}\frac{r^2dr}{\sqrt{B(r)}}\label{Vmax}
 \end{eqnarray}
where $r_{+}$ is horizon, and $r_{\infty}$ is an IR cutoff. 
Now we can use the Poincare coordinate $w=\frac{r}{r_{+}}$  to evaluate
integral (\ref{Vmax}) as
\begin{eqnarray}
&&V(\gamma_{max})=r_{+}^3\int_{\epsilon}^{1}
\frac{dw}{w^4\sqrt{b(w)}}
\label{Vmax-z}
 \end{eqnarray}
where $\epsilon\to 0$ is an $UV$ cutoff. We also have  
\begin{eqnarray}&&
b(w)= b_1w^{1+z/2} -\frac{b_{-2}}{w^2}.
 \end{eqnarray}
It may be noted as  
$z=-\frac{4\tilde Q^2\xi}{\Lambda+\tilde Q^2\xi}\geq 3,
\Lambda=-\frac{3}{L^2}$, so  $z=4$ is an interesting solution.  
 In this case, the coefficients $b_n$ are given as following
\begin{eqnarray}
 &&
b_1= \Big(\frac{\Lambda+\tilde Q^2\xi}{5}\Big)r_{+}^2-\frac{1}{3}\left[\frac{\tilde V_0}{2}
\right],\\&&
b_{-2}=\frac{r_{+}^2(\Lambda+\tilde Q^2\xi)}{5}.
\end{eqnarray}

Now  we obtain 
\begin{eqnarray}\label{volume22}
&&V(\gamma_{max})=r_{+}^3\Big(\frac{A}{3840
   (-b_{-2})^{9/2}}
\Big)+\frac{r_{+}^3}{2 \epsilon ^2 \sqrt{-b_{-2}}}
\end{eqnarray}
here $A = 640 b_{-2}^3 (b_1-3 b_{-2})$. 
Here the bulk charge $\tilde{Q}$ is dual  to 
the magnetic charge (strength) $ H$ of  the  boundary theory, and  
$H$  varying smoothly. So, we can 
 the volume (\ref{volume22}) for $\mathcal{O}(\frac{1}{H^3})$, and  obtain,
\begin{eqnarray}
&& F_{DAdS}=\frac{\sqrt{5} r_{+}^2\sqrt{-\xi } \left(2 L^2 \xi  \tilde Q+3\right)}{48 \pi  G
   L^3 \xi ^2 \tilde Q^3}-\frac{\sqrt{5} r_{+}^2\sqrt{-\xi } \left(2 L^2 \xi  \tilde Q^2+3\right)}{32 \pi  G
   L^3 \xi ^2 \tilde Q^3 r^2 \epsilon ^2}.
 \end{eqnarray}
 Now this equation contains both the finite and  divergent parts of the holographic fidelity susceptibility.
 However, we can regularize it by subtracting it from the background AdS geometry, 
 and obtain the regularized finite fidelity susceptibility. 
 \begin{eqnarray}
\Xi_F  &=& F_{DLADS}- F_{LAdS}\nonumber \\ &=& \frac{\sqrt{5} r_{+}^2\sqrt{-\xi } \left(2 L^2 \xi  \tilde Q+3\right)}{48 \pi  G
   L^3 \xi ^2 \tilde Q^3}. \label{fid-holo}
 \end{eqnarray} 
It may be noted that  that the regularized fidelity susceptibility calculated holographically in  (\ref{fid-holo})
is same as the fidelity susceptibility of the boundary theory obtained in (\ref{XIF}). 
So, we can write 
\begin{eqnarray}
&&\frac{\sqrt{5} r_{+}^2\sqrt{-\xi } \left(2 L^2 \xi  \tilde Q+3\right)}{48 \pi  G
   L^3 \xi ^2 \tilde Q^3}=\frac{N}{8 q H^3}(qH+4\beta^2)
 \end{eqnarray}
Now if we  holographically identify $\tilde Q=H$, 
we can also identify many  parameters in the bulk to boundary theories. 
In fact, from this identification, we obtain 
\begin{eqnarray}
&&\frac{2 \sqrt{5}L^2 \xi  r_{+}^2 \sqrt{-\xi} }{48 \pi  G
   L^3 \xi ^2 }\equiv\frac{ N}{8  },\\&&
\frac{3\sqrt{5} r_{+}^2 \sqrt{-\xi}}{48 \pi  G
   L^3 \xi ^2 }=\frac{N\beta^2}{2 q }.
\end{eqnarray}
So, the  number of boundary quantum systems $N$ and $\beta^2 q^{-1}$
  can be expressed as 
\begin{eqnarray}
&&N^{\mbox{boundary}}=\Big(-\frac{16 \sqrt{5}L^2  r^2_{+}  }{48 \pi  G
   L^3 \sqrt{-\xi} }\Big)^{\mbox{Bulk}},\\&&
\Big(\frac{\beta^2}{ q }\Big)^{\mbox{boundary}}=\Big(\frac{3}{8L^2 \xi}\Big)^{\mbox{Bulk}}.
\end{eqnarray}

It may be noted  that $\xi<0$, for this system.  
So,  we have analyzed a system of  bosonic particles in a 
magnetic field, and we obtained the fidelity susceptibility for this system. As this system 
was a non-relativistic system, it was expected to be dual to a AdS-Lifshitz spacetime. 
We have demonstrated that this theory is dual to a 
Einstein-Dilaton-Maxwell-AdS-Lifshitz, and the fidelity susceptibility calculated 
from the bulk using this theory matches with the fidelity susceptibility of the boundary 
theory. 

\section{Conclusion}

In this paper, we propose that the fidelity susceptibility of a non-relativistic system can be obtained holographically 
from the holographic complexity of a Lifshitz-AdS theory. We use this proposal to holographically analyze the 
 fidelity susceptibility of non-relativistic field theories, and demonstrated that fidelity susceptibility of the bulk theory 
 is the same as the boundary theory.   So, using a 
Einstein-Dilaton-Maxwell-AdS-Lifshitz theory, we explicitly 
  demonstrated that the 
 fidelity susceptibility obtained from this  
  bulk geometry is equal to the   fidelity susceptibility of a  
  bosonic many-body system. 

It may be noted that as the  boundary system considered explicitly in this paper 
described a simple system, it was possible to calculate the 
fidelity susceptibility  for this system, both in the boundary and in the bulk. 
However, it is not always possible to calculate the fidelity susceptibility for the boundary 
and the bulk system. It would be difficult to analyze the strongly coupled field theories, 
and perform such calculations in the field theory side of the duality. However, it is known that 
a strongly coupled limit of the  field theory can be holographically analyzed using a
weakly coupled limit on the gravity side of the duality. Thus, for such  non-relativistic 
systems, where such calculations cannot be performed on the field theory side of the duality, 
this holographic calculations can be performed using the gravitational side of the duality. 
As fidelity susceptibility is an very important quantity in condensed matter systems, and many 
condensed matter systems can be modeled using conformal field theories, it would be possible 
to use the formalism developed in this paper for analyzing such condensed matter systems. 

It  is possible to describe condensed matter systems like Weyl semi-metal as strongly coupled 
systems \cite{wely}, it would be possible and interesting to use the results of this paper for holographically 
analyzing such fidelity susceptibility of such system. 
As we have generalized the fidelity susceptibility to Lifshitz geometries, and 
Lifshitz geometries can have important condensed matter applications 
\cite{cond1, cond2, cond4, cond5, cond6}, 
the results of this paper can have interesting condensed matter applications. 
So, it would be   interesting to analyze  realistic 
condensed matter systems, and then use the Lifshitz holography to understand 
the behavior of fidelity susceptibility for such condensed matter systems. 
It is also possible to study various interesting time-dependent 
generalizations of this solution. It is expected that such a  time-dependent 
system on the boundary will be dual to some time-dependent Lifshitz bulk solution. 
It may be noted that fidelity susceptibility for time-dependent geometries has been studied
\cite{time}, and it is expected that this formalism can be generalized to bulk geometries 
with Lifshitz symmetry.  This would be interesting as such time-dependent 
systems are important in condensed matter physics \cite{time1}. It would be interesting to analyze such geometries, 
and use them 
to understand the boundary fidelity susceptibility. Thus, the proposal developed in this paper can be used to analyze 
interesting condensed matter systems, and calculate the fidelity susceptibility for such systems.  
We would like to point out that it would be interesting to extend  this work to other geometries \cite{geom1, geom2}, 
and calculate the 
fidelity susceptibility for such geometries.

\end{document}